\documentclass[superscriptaddress,
amsmath,amssymb,
]{revtex4-2}

\usepackage{units,tabularx,graphicx,dcolumn,bm,tikz,soul,geometry,hyperref,braket}

\renewcommand{\vec}[1]{\bm #1}
\newcommand{\bx}{\bm{x}}
\newcommand{\bk}{\bm{k}}
\newcommand{\dd}{\textrm{d}}
\newcommand{\aaa}{\textsc{a}}
\newcommand{\bbb}{\textsc{b}}
\newcommand{\iii}{\textsc{i}}
\newcommand{\jjj}{\textsc{j}}
\newcommand{\ii}{\textrm{i}}
\newcommand{\mc}[1]{\mathcal{#1}}

\newcommand{\squeezeup}{\vspace{-5.7mm}}

\begin{document}

\title{Vacuum entanglement probes for ultra-cold atom systems}

\author{Cisco Gooding}
\affiliation{School of Mathematical Sciences, University of Nottingham, University Park, Nottingham, NG7 2RD, UK}

\author{Allison Sachs}
\affiliation{Department of Physics and Astronomy, University of Waterloo}

\author{Robert B. Mann}
\affiliation{Department of Physics and Astronomy, University of Waterloo}
\affiliation{Perimeter Institute for Theoretical Physics}

\author{Silke Weinfurtner}
\affiliation{School of Mathematical Sciences, University of Nottingham, University Park, Nottingham, NG7 2RD, UK}
\affiliation{Centre for the Mathematics and Theoretical Physics of Quantum Non-Equilibrium Systems, University of Nottingham, Nottingham, NG7 2RD, UK}

\date{\today}

\begin{abstract}

 This study explores the transfer of nonclassical correlations from an ultra-cold atom system to a pair of pulsed laser beams. Through nondestructive local probe measurements, we introduce an alternative to destructive techniques for mapping BEC entanglement. Operating at ultralow temperatures, BEC density fluctuations emulate a relativistic vacuum field. We show that lasers can serve as Unruh-DeWitt detectors for vacuum BEC phonons. A quantum vacuum holds intrinsic entanglement, transferable to distant probes briefly interacting with it - a phenomenon termed `entanglement harvesting'. Our study accomplishes two primary objectives: first, establishing a mathematical connection between a pair of pulsed laser probes interacting with an effective relativistic field and the entanglement harvesting protocol; and second, to closely examine the potential and persisting obstacles for realising this protocol in an ultra-cold atom experiment.
\end{abstract}

\maketitle

\section{Introduction}
\begin{figure}[ht]
    \centering
    \includegraphics[width=0.85\columnwidth]{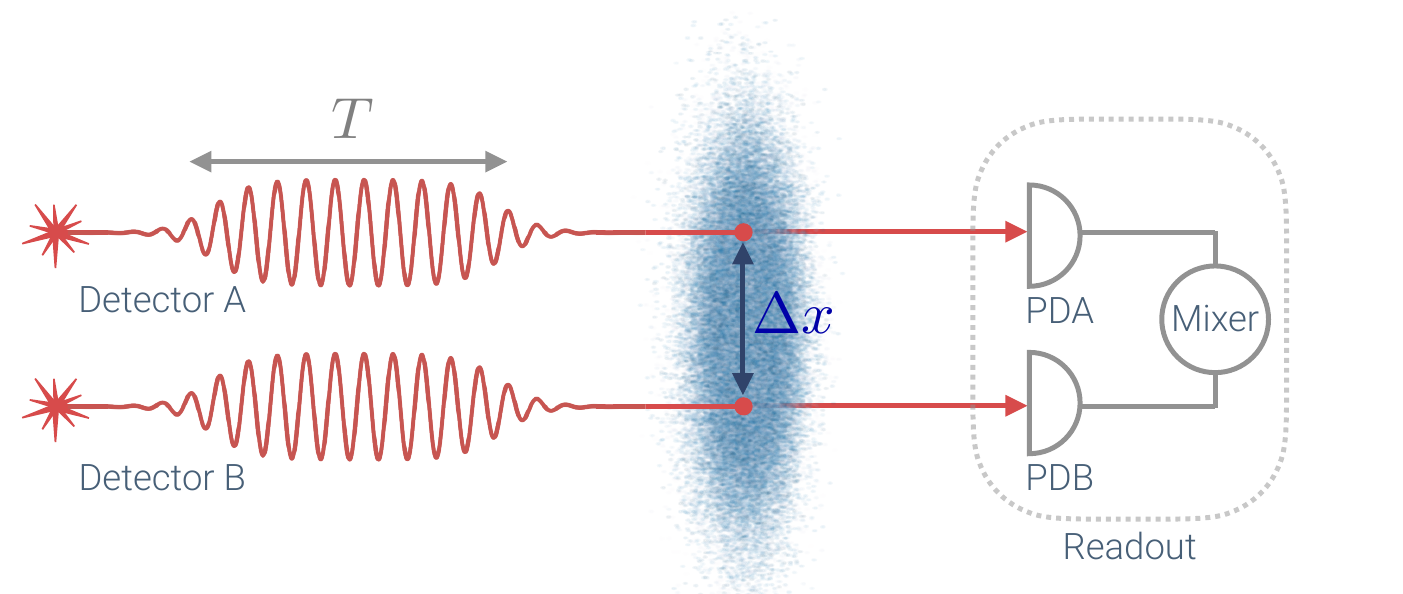}
    \caption{Schematic representation of the proposed experiment. The setup utilizes a pair of pulsed laser beams as local detectors of density fluctuations in a Bose-Einstein condensate. The pulses become entangled by the BEC, and enter photodiodes for readout.}
    \label{fig:diag}
    \squeezeup
\end{figure}

The quantum vacuum has long been known to contain entanglement~\cite{salton2015aa, summers87aa}. It has been predicted that this entanglement can be transferred to two or more spatially separated local probes briefly interacting with the vacuum state~\cite{reznik05aa}. Such entanglement extraction even persists for causally disconnected probes~\cite{pozas15aa}. The process of \textit{entanglement harvesting} is theoretically well-understood, but an experimental implementation remains unrealized, due in part to the level of idealisation. We propose here that probing such fundamental properties of vacuum states, as for example provided by an ultra-cold atom system such as a Bose-Einstein Condensate (BEC), is within experimental reach. Concretely, we show that two laser pulses simultaneously interacting with an oblate BEC at two distinct locations can become entangled, as depicted in Figure~\ref{fig:diag}. As we consider continuous, nondestructive measurement of a BEC by local probes, our approach can be considered complementary to existing destructive measurements aimed at mapping out the global entanglement structure of BECs \cite{PhysRevLett.128.020402,doi:10.1126/science.aao2254,Hamley2012,PhysRevA.65.033619,doi:10.1126/science.aau4963}.

The crucial parameters that determine entanglement are the separation of the detectors, their effective detector gaps, and their interaction time with the quantum field. For the lasers serving as detectors in our proposed setup, the interaction time is determined by the pulse duration. Utilising negativity as a measure, we find that for a given effective gap $\Omega$,
the extracted entanglement is maximal for interaction times that are shorter than $\sim 2\pi/\Omega$. This implies that optimizing the harvest does not require spectral resolution. 
We will elaborate on the connection between the effective detector gap and properties of the laser probes in what follows.

A regime is identified for which the accuracy of the relativistic analogy is high, while also yielding a negativity that is maximal for experimentally
accessible parameters. We then discuss how an inseparability condition could allow harvested entanglement to be verified, without the state reconstruction that one must often perform to access
the negativity. Essentially, we propose an alternative pathway for extracting entanglement
from the vacuum state simulated in an ultracold atom system. The vacuum field provided by a BEC extends the single-mode vacuum simulated in nonlinear crystals \cite{Settembrini2022,EMHarvest2023}; at the same time, our experiment is a natural generalisation of \cite{entangledmodes} for pulsed laser probes, arbitrarily displaced from one another.

\section{Physical System}
The laser-BEC system  we consider can be modelled by the Lagrangian $L_{tot}=L_{BEC}+L_{EM}+L_{int}$, which describes the BEC and electromagnetic (EM) fields ($L_{BEC}$ and $L_{EM}$, respectively), along with their interaction. The homogeneous quasi-2d BEC Lagrangian is given by $L_{BEC}=\int d\vec{x}\, \mathcal{L}_{BEC}$, where
\begin{align}\label{Lbec}
\mathcal{L}_{BEC}=i\hbar\Phi\partial_t \Phi^*+\frac{\hbar^2}{2m}|\nabla\Phi|^2+\frac{g_{2d}}{2}|\Phi|^4\,
\end{align}
$\Phi=\Phi(t,\vec{x})$ is the BEC field, $m$ is the atomic mass, $g_{2d}$ is the effective 2d coupling constant and $\vec{x}$ is a point on the $(x,y)$ plane~\cite{Pethick2008,gooding20aa}.
A single BEC can without loss of generality be described by a constant real $\Phi = \Phi_0$. 
 Fluctuations $\delta \Phi$ about 
 this background correspond to variations in the  BEC density:  $\hat{\phi}\equiv \Phi_0(\delta \Phi+\delta \Phi^\dagger)$ \cite{Unruh2022}.

The EM field includes two lasers propagating in the $z$ direction, which can be described in the Coulomb gauge (i.e. $E=\partial_t A$) by the free Lagrangian
\begin{align}
    L_{EM}=\frac{1}{2}\sum_{\iii=\aaa,\bbb}\int dz\, \left((\partial_t A_\iii)^2-(\partial_z A_\iii)^2\right)\, ,
\end{align}
such that $A_\iii=A_\iii(t,z)$ and $\iii=\aaa, \bbb$ for the two laser probes. For local interactions at the points of intersection $\vec{x}_\aaa$ and $\vec{x}_\bbb$ between the lasers and the BEC plane ($z=0$), the interaction Lagrangian is given by
\begin{align}\label{Lintj}
    L_{int}=\alpha \sum_{\iii=\aaa,\bbb}E_\iii(t,0)^2 |\Phi(t,\vec{x}_\iii)|^2 \, ,
\end{align}
where $E_\iii=\partial_t A_\iii$ and $\alpha$ is the atomic polarizability. See~\cite{gooding20aa} for further details about the single-probe version of this model, and the interpretation of the linearized theory in terms of an effective relativistic field for BEC density fluctuations. Note that the $(2+1)$-dimensional nature of the BEC field $\Phi$ and the $(1+1)$ nature of the probe fields $E_\ii$ in the geometry depicted in Figure~\ref{fig:diag} leads to effective pointlike interactions between the probe lasers and the BEC field, representing local detectors for BEC phonons. 

For the purpose of extracting entanglement, it is convenient to use the interaction picture.
To this end,
the Hamiltonian associated with 
\eqref{Lintj} is 
$\hat{H}_\iii(t)=-\alpha (\hat{E}_\iii(t,0))^2\, |\Phi(t,\bx_\iii)|^2$, which becomes
\begin{align}\label{Hint}
     \hat{H}_\iii(t)=-\alpha E_{\iii0}(t,0)\delta \hat{E}_\iii(t,0)\hat{\phi}(t,\bx_\iii)\, ,
\end{align}
for linear fluctuations $\delta\hat{E}_\iii(t,z)$ about the coherent amplitudes $E_{\iii0}(t,z)$. These coherent amplitudes determine the laser pulse shape and duration, while the EM fluctuation operators $\delta\hat{E}_\iii(t,z)$ serve as local detectors for the BEC density fluctuation field $\hat{\phi}$.
 
In the absence of coupling to the EM field, density fluctuations represented by 
 $\hat{\phi}$ obey
a wave equation of the form
\begin{align} \label{GWE}
    \partial_t^2 \hat\phi(t,\bx)  
    +
    G(-\ii \nabla )\hat\phi(t,\bx)
    =0 \, ,
\end{align}
which has the associated 
dispersion relation $\omega_{\bk }^2=G(\bk )$. In particular, low-energy excitations in homogeneous and time-independent systems are often governed by
\begin{equation}
    \omega_k^2=c^2k^2 \pm \epsilon^2 k^4 \, ,
    \label{recoverLinearDispersion}
\end{equation}
where $\epsilon$ is an expansion parameter quantifying nonlinear dispersive corrections to the effective relativistic behaviour exhibited by low-$k$ modes, which propagate at speed $c$. The sign choice ($\pm$) characterizes the type of dispersion: $(+)$ implies Bogoliubov dispersion, as obeyed by linearized phonons of the BEC Lagrangian~\eqref{Lbec}, while $(-)$ describes subsonic dispersion, as obeyed by interface waves on (super)fluids. We will restrict our attention to BECs, and work in units with $c=\hbar=1$ (emphasizing that the natural units involve setting the phonon propagation speed $c$ to unity, not the speed of light), unless otherwise indicated.


\section{Unruh-DeWitt Detectors}
To adapt the entanglement harvesting protocol to systems with non-trivial dispersion relations, we consider a pair of detectors using the Unruh-DeWitt (UDW) model~\cite{unruh76aa,dewitt79aa}. We shall later relate these simple detectors to a pair of local laser pulses, which probe the BEC via \eqref{Hint}. The UDW detectors couple to the field $\hat \phi(t,\bx)$ according to the interaction Hamiltonian 
$\hat H(t)=\hat H_{\aaa}(t)+\hat H_{\bbb}(t)$, with
\begin{align}\label{UDW-Ham}
    \hat H_{\iii}(t)=\lambda(t)
        \hat m_{\iii}(t)
        \int \dd^2 \bx \,
        F_{\iii}\left(\bx\right)
        \hat \phi(t,\bx),
\end{align}
where $\iii = \aaa,\bbb$ represents detectors $A$ and $B$ and where $\hat \phi$ in general
obeys \eqref{GWE}; 
the full Hamiltonian is given in the supplementary material. The factor $\lambda(t)=\lambda\,\chi\left(\frac{t}{T}\right)\leq \lambda$ includes a time-independent coupling $\lambda$ and a switching function $\chi\left(\frac{t}{T}\right)$, which combines with the smearing functions $F_{\iii}(\bx)=F\left(\frac{\bx-\bx_{\iii}}{\sigma}\right)$ to define the region of $(2+1)$ spacetime in which each detector interacts with the field. The detector positions $\bx_\iii$ correspond to the centroids of the intersections of the laser probes with the BEC plane, and $\sigma$ characterizes their spot size. The detectors 
both couple to the field
for a duration $T$, given in the physical system by the temporal widths of the laser pulses.
Assuming laser profiles without angular dependence, the transverse spatial extent of the probes can be modelled by a rotationally symmetric smearing function, $F(\bx)=F(|\bx|)$
(whose Fourier transform is
$\widetilde{F}[|\bk|]$), 
with $|F(|\bx|/\sigma)|$ normalized to unity when integrated with respect to $\bx$. Each detector couples to the field through the operator $\hat m_{\iii}(t)$, given in the two-level case by the simple monopole operator
\begin{align}
    \hat m_{\iii}(t)=e^{\ii\Omega t}\sigma_{\iii}^++e^{-\ii\Omega t}\sigma_{\iii}^-~,
    \label{UDW-mon}
\end{align}
where $\Omega$ is the detector energy gap, and $\sigma_{\iii}^\pm$ are Pauli ladder operators for each detector.

From \eqref{UDW-Ham}, the time evolution operator of the system in the interaction picture 
is $\exp[-i \int dt H(t)]$.
The effect of the interaction on the detectors is determined by the reduced density operator $\hat \rho_{\aaa\bbb}=\textrm{Tr}_\phi(\hat \rho_T)$, obtained by taking the trace with respect to $\hat \phi$. The leading-order contributions to $\hat \rho_{\aaa\bbb}$, for initially uncorrelated UDW detectors weakly coupled to a dispersive vacuum field $\hat \phi$, are derived in the Appendices.

\section{Entanglement Harvesting Protocol}
Entanglement harvesting is the process of transferring entanglement from a quantum field to a pair of detectors. The associated protocol provides a way of probing the entanglement structure of a quantum field \cite{brown13aa}, as well as geometric and topological aspects of spacetime \cite{martinez16aa,ng18ab}. By suitable repetition of the harvesting protocol, field entanglement could potentially be extracted in a sustainable way~\cite{martinez13aa}, providing a valuable resource for quantum computation.

In the standard harvesting protocol, a pair of (two-level) UDW detectors interact in a given region of spacetime with a quantum field in a known state (e.g. vacuum~\cite{pozas15aa}, Fock~\cite{Tjoa2020}, thermal~\cite{simidzija18}, or coherent~\cite{simidzija17}). After the interaction, any ensuing correlations between the detectors can be obtained from the associated reduced density operator. The amount of harvested entanglement is quantified using the negativity, which, in this case, is proportional to the concurrence~\cite{martinez16aa}. Both negativity and concurrence serve as entanglement measures for pairs of two-level systems, even for mixed states~\cite{horodecki09}.

For identical detectors, the negativity can be expressed in terms of the vacuum excitation probability of each detector, $\mc L$, and a non-local term $\mc M $ that can be interpreted as the probability for virtual particle exchange between the detectors \cite{reznik05aa}. Concretely, the negativity is given by
\begin{equation}
	\mc N =\max\left[|\mc M |-\mc L, 0\right] +\mathcal{O}	(\lambda^3)\, , \label{neg}
\end{equation}
which reflects how entanglement emerges as a competition between the (non-local) virtual exchange term $\mc M $ and the noise associated with the vacuum excitation probability $\mc L $ \cite{pozas15aa}.

Using the density matrix elements derived in the Appendices, taking $\chi(t)$ to be Gaussian for convenience, one finds that $\mc M$ and $\mc L$ can be expressed as 
\begin{align} \label{LandMint}
    	\begin{pmatrix}
    	    \mathcal{L} \\
         \mathcal{M}\\    	\end{pmatrix}=\frac{\lambda^2 T^2}{4}\int_0^\infty 
                    \dd |\bk|\, \mc F(|\bk|)\,
 \cdot\begin{pmatrix}
    G_{\mathcal{L}}\\
    G_{\mathcal{M}}\\
    	\end{pmatrix}
    	\, ,	
\end{align}
where 
$\mc F(|\bk|)= |\bk|
(\widetilde{F}[|\bk|\sigma])^2 /2\omega_{\bk }$ and
\begin{align}
    G_{\mc L}(\bk) = \exp(-T^2 (\Omega+\omega_{\bk })^2/4)\, ,\,
    G_{\mc M}(\bk)=\mathcal{Q}(\bk)\exp(-T^2(\Omega^2+\omega_{\bk }^2)/4)
\end{align}
with $\mathcal{Q}(\bk)\equiv J_{0}(|\bk|\,\Delta \bx) (i\text{erfi}\left(T\omega_{\bk}/2)\right)-1)$, 
 $\Delta \bx\equiv |\bx_\aaa-\bx_\bbb|$, and $\text{erfi}(x)\equiv -i\,\text{erf}(ix)$. The dependence of the negativity estimator $|\mc M |-\mc L$ with
relative scaling of the detector size, energy gap, and mode frequency is now apparent. For wavevectors much greater than $1/4\Delta \bx$, the Bessel function oscillates as $\sqrt{\frac{2}{\pi |\bk| \Delta x}}\cos(|\bk| \Delta x-\pi/4)$, which strongly suppresses these contributions to the integral. 

Dispersion becomes relevant when the quartic term in ~\eqref{recoverLinearDispersion} becomes comparable to the quadratic term; their point of equivalence is the crossover scale $k_c = 1/\epsilon$, indicating a transition from the linear (phononic) band to the dispersive regime. If $k_c$ is sufficiently larger than the inverse detector size $k_0\sim 1/\sigma$, high-$|\bk|$ contributions of the integrands in ~\eqref{LandMint} are suppressed; therefore, for a detector with size sufficiently larger than $1/k_c$, the effects of dispersion for harvesting are negligible.

\begin{figure}[ht]
    \centering
     \includegraphics[width=0.8\linewidth]{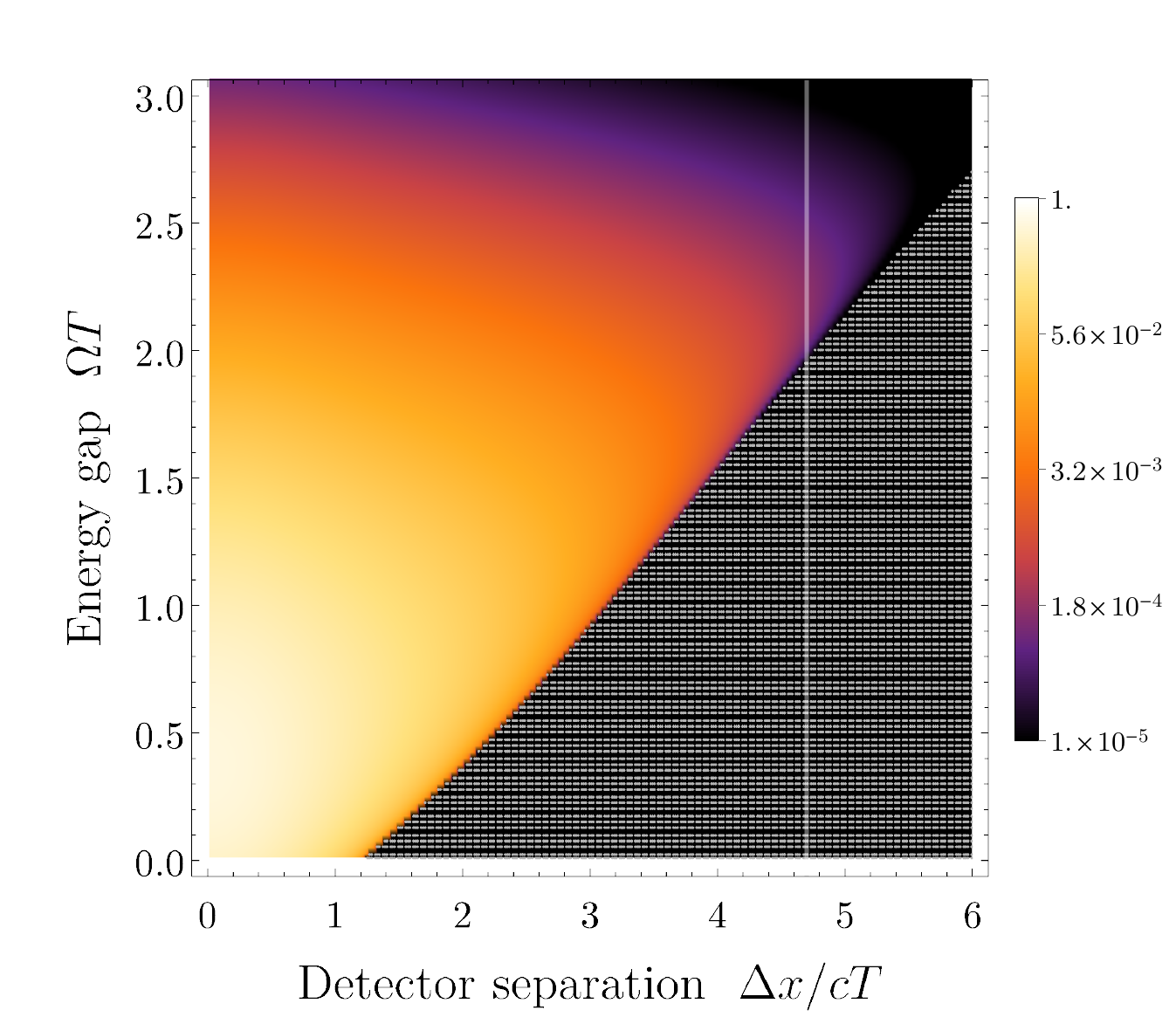} 
    \caption{Logarithmic Concurrence (i.e. logarithm of twice the negativity) as a function of the gap $\Omega T$ and the separation $\Delta x/cT$ with a UV cutoff determined by the detector size;  concurrence is in units of $\lambda^2 T^2$, and is zero in the right-hand dotted region. The vertical line
    denotes the communication boundary, given the sound speed and dispersion; all points to the right of this line represent causally disconnected detectors.}
    \label{fig:conc}
    \squeezeup
\end{figure}
\vskip 0.3 cm

Figure~\ref{fig:conc} shows the behaviour of the logarithmic negativity as a function of the detector separation $\Delta x$ and the energy gap $\Omega$, applicable to a BEC being probed by a highly-focused laser with spot size $\sigma= 3\mu\text{m}$. The energy gap is multiplied by the (fixed) interaction time, $T= 3\text{ms}$, which represents the pulse duration and was chosen to ensure the negativity is optimal for accessible BEC frequencies. The separation is scaled by the (also fixed) sound travel distance $c T$, with $c$ chosen to 
be achievable for a Rubidium BEC, as discussed below. It is immediately apparent that harvested entanglement is appreciable only for $\Omega T\sim 1$. The vertical line separates the region where the detectors can signal one another (left) from the region where the detectors maintain spacelike separation throughout the interaction (right). The spacelike region is determined by the criteria
\begin{align}\label{spacelike}
    \Delta x\ge 4 c T +2\sigma\, ,
\end{align}
which is chosen so that no causal contact can be made between times $-2 T$ to $2T$, with Gaussian peaks separated by at least $2\sigma$. For a thorough investigation of how to distinguish causal and non-causal contributions in entanglement harvesting, see \cite{PhysRevD.104.125005}.

For an accurate depiction of the laser-BEC interaction, a refinement of the standard harvesting analysis is required. This involves replacing the UDW detectors with continuous-detector fields, to describe the laser probes used to extract entanglement from the BEC vacuum field. Continuous-detector fields can be conceptualised as an infinite array of harmonic oscillator detectors, each characterised by a distinct detector gap~\cite{gooding20aa}. A single pair of weakly-coupled harmonic detectors prepared in their ground states can be well approximated with the two-level model~\cite{PhysRevD.101.085017}; in particular, the leading order oscillator negativity is identical \cite{PhysRevD.79.085020}. 

Evidently, the simple case of vacuum harvesting with two-level detectors serves not only as an intuitive guide to understand more complex harvesting scenarios, but also as a quantitative guide, for a nontrivial class of weakly-coupled detectors. To clarify the extent to which this applies to the laser-BEC system, some caveats are in order. Strictly speaking, the continuous-detector fields corresponding to our pair of pulsed laser probes can indeed be decomposed into infinitely-many harmonic oscillator detectors, and each pair of individual harmonic modes shared by the laser probes can be described (in the weakly-coupled limit) by the two-level results shown above. However, that does not imply that it is a straightforward matter to isolate such a pair of harmonic modes from the laser probes after interaction, in hopes of verifying the predicted negativity. Given a laser pulse of duration $T$, we learned from Figure~\ref{fig:conc} that harvesting is negligible outside of a small frequency band around $\Omega T\sim 1$, which is a sub-cycle  regime; hence, the pulse duration is insufficient to allow individual harmonic modes to be reliably distinguished. Fortunately, due to the small size of the frequency band in question, the pulsed laser probes behave as if they were monochromatic, with the corresponding detector gap taking the (approximately constant) value $\Omega\sim 1/T$.

Restricting attention to local probes (i.e. vanishing spatial extent of the smearing function $F$), the form of~\eqref{Hint} indicates that the coherent amplitudes 
$E_{\iii0}(t,0)$
serve as switching functions, enabling precise control of the interaction time. The specific choice of Gaussian switching has been shown to work well for harvesting~\cite{pozas15aa}. In the Appendices we demonstrate that for a fixed detector gap, the negativity for a pair of local laser probes is determined by reduced density matrix elements of the same two-level UDW form. The same reasoning used in the two-level case then leads to the conclusion that each pair of identical modes from the two local probes are entangled, with a negativity described by the pointlike limit of~\eqref{neg}. The argument given in the preceding paragraph implies that such a negativity prediction can also be used to estimate entanglement of the entire pulse pair. Accuracy in this estimate could be improved by selecting an entanglement measure that more naturally incorporates time-dependence, as we discuss in the following section. Regardless of the choice of measure, the amount of predicted entanglement can be systematically optimised (see techniques in \cite{PhysRevA.67.042314}, for instance), reducing ambiguity in experimental parameter selection.

\section{Experimental Outlook}
We now consider specific experimental conditions for realising the entanglement harvesting protocol in an oblate BEC, using laser pulses as effective vacuum probes. The in-plane spatial extent of the BEC is taken to be roughly $l_{\text{BEC}}=10^{-4}\text{m}$. We further assume the BEC is composed of Rubidium: this fixes the healing length $\xi=6.31\times10^{-8}\text{m}$ and the sound speed $c=8$ mm/s (restoring SI units) \cite{verhelst17}. The corresponding dispersive crossover scale follows from $\xi=\sqrt{2}\epsilon /c$. Since the healing length is nearly two orders of magnitude smaller than the detector size, $\sigma\approx 3\mu\text{m} \approx 50\, \xi$, the detectors are only sensitive to linearly-dispersive (phononic) modes, and the relativistic analogy is expected to be accurate. The low sound speed also facilitates spacelike harvesting by relaxing prohibitively tight constraints on the pulse duration, which is set by the (fixed) interaction time to be $T= 3\text{ms}$. The signal for the logarithmic negativity (illustrated in Figure~\ref{fig:conc}) may be optimised further over BEC species to maximise BEC lifetime, which in turn allows for higher interaction times and correspondingly higher logarithmic negativities. 

The initial beams are each prepared with two modulation sidebands, symmetrically arranged on opposite sides of an atomic resonance, to balance Stark potentials, minimising disturbance to the $2\text{d}$ BEC. This technique was used previously in a single-beam setting \cite{gooding20aa}, and we elaborate on its use for harvesting in the Appendices. Beating caused by the modulation sidebands in the laser converts phase fluctuations into amplitude fluctuations, and the coherent components of the laser fields amplify quantum fluctuations. As a result, the readout stage shown in Figure~\ref{fig:diag} can be achieved for the modulated laser probes using multi-tone heterodyne scheme, such as dual-arm heterodyne detection (i.e. splitting the modulation bands with a dichroic mirror and heterodyning each band); details of signal analysis for such a readout scheme are the subject of separate work. Upon demodulation, the relevant photocurrent output for each beam corresponds to a dark-port signal, with the common-mode coherent amplitude contribution filtered out.

Photon statistics extracted from the photocurrents strongly depend on the reduced density operator $\hat{\rho}_{\aaa\bbb}$, which characterizes how variances in the laser quadratures are affected by the BEC interaction~\eqref{Hint}. In parallel with heterodyne detection schemes, the readout scheme described here yields variances of quantum optical modes entering each photodetector. 

From an operational point of view, while it is possible to directly measure the negativity by completely reconstructing the reduced state of the laser pair, the quantum metrology would be nontrivial. For continuous-mode lasers, it is simpler to proceed indirectly, using a more accessible entanglement measure: the \textit{inseparability}, $\mc I$ \cite{entangledmodes}. Conceptually, the negativity provides a lower bound for the inseparability with respect to all local linear unitary Bogoliubov transformations between pairs of joint detector quadratures \cite{entangledmodes}; based on the DGCZ condition \cite{PhysRevLett.84.2722}, the inseparability $\mc I$ is less than unity unless the detectors are separable. The ``inseparability condition'' can then be expressed as
\begin{align}\label{DGCZ}
\mc I(\Omega) = V(\hat{q}_+(\Omega)) + V(\hat{p}_-(\Omega))<1\, 
\end{align}
where $V$ denotes the variance and $\hat{q}_+(\Omega)$, $\hat{p}_-(\Omega)$ denote the joint amplitude-sum quadrature and the phase-difference quadrature of the two laser probes (respectively), normalised such that vacuum variances sum to unity.

A suitable generalisation of (\ref{DGCZ}) has utility for experimentally verifying the entanglement encoded in the reduced density operator of our pulsed detectors. To qualify, a generalised inseparability should be bounded by the corresponding entanglement measure. To be useful, it should be possible to approximately saturate the bound for some set of local detector field configurations. 
In the Appendices, we discuss a possible generalised inseparability, and specify how it can be applied to our experimental setup. 
The connection between spectral mode measurements and negativity has only been explicitly demonstrated for stationary laser signals \cite{Zippilli_2015}; establishing this connection for general modulated detector pulses is beyond the scope of this paper, though there are indications that the pulsed case behaves similarly~\cite{Zhang2015,Wenger2004}. We also note that a closely-related time-domain entanglement condition has been used to verify EPR correlations in optical fibers~\cite{PhysRevLett.86.4267}. While further work is required to rigorously demonstrate feasibility, measurements of photon correlations in optomechanical systems suggest that our proposal is within experimental reach \cite{Settembrini2022,entangledmodes}. 

\section{Conclusion}
The connection between a pair of modulated laser pulses coupled to an ultra-cold atom system with an idealised pair of Unruh-DeWitt detectors probing a relativistic quantum vacuum establishes a synthetic quantum system to implement the entanglement harvesting protocol. A key step towards this was to demonstrate that the effects of dispersion can be neglected, through judicious choice of experimentally accessible parameters. In turn, the proposed scheme has potential to serve as a local, unequal-time alternative to absorption-based imaging techniques to probe continuous variable entanglement in non-equilibrium ultra-cold atom systems. 

\section*{Acknowledgements}
The authors thank August Geelmuyden for figure generation, and Erickson Tjoa, Tales Rick Perche, and Eduardo Martin-Martinez for discussions at various stages of this project.
This work was supported in part by the Natural Sciences and Engineering Research Council of Canada. We acknowledge support provided by the Leverhulme Research Leadership Award (RL-2019- 020),
the Royal Society University Research Fellowship
(UF12011) and the Royal Society Enhancements Awards and Grants
(RGF\textbackslash EA\textbackslash 180286, RGF\textbackslash EA\textbackslash 181015,  RPG\textbackslash 2016\textbackslash 186), and partial
support by the Science and Technology Facilities Council (Theory Consolidated Grant ST/P000703/1), the Science and Technology Facilities Council on Quantum Simulators for Fundamental Physics (ST/T006900/1) as part
of the UKRI Quantum Technologies for Fundamental
Physics programme. For the purpose of open access, the authors have applied a CC BY public copyright licence to any Author Accepted Manuscript version arising.

\section*{Appendix I: Calculation of the reduced density matrix}

\subsection{Two-level Unruh-DeWitt Detectors}
\label{ap:calculation-reduced}

The UDW model consists of
the   Hamiltonian
\begin{align}
  H =  \frac{1}{2}\int d^2x \left(\pi^2(x)
  + | \tilde{G}(-\ii \nabla) \phi|^2
  \right) + \sum_{\iii = \aaa,\bbb} \Omega_{\iii} \ket{1_\iii}\bra{1_\iii}  + H_{int}
\end{align}
of a scalar field $\phi$ and
two 2-level detectors $(\aaa,\bbb)$ having 
ground $\ket{0_\iii}$
and excited states $\ket{1_\iii}$ separated
by
respective energy gaps $(\Omega_\aaa,\Omega_\bbb)$. 
Here  $\tilde{G^\dagger}\tilde{G} = {G}(-\ii \nabla)$ is an arbitrary function of the $\nabla$-operator, where $\omega_{\bk }^2=G(\bk )$ is the general dispersion relation.

The interaction between the detectors and the field is
\begin{align}\label{UDW-Ham}
    H_{int}=
    \sum_{\iii = \aaa,\bbb}
    \lambda_\iii(t)
        \hat m_{\iii}(t)
        \int \dd^2 \bx \,
        F_{\iii}\left(\bx\right)
        \hat \phi(t,\bx),
\end{align}
where  
\begin{align}
    \hat m_{\iii}(t)=e^{\ii\Omega_\iii t}\sigma_{\iii}^+ + e^{-\ii\Omega_\iii t}\sigma_{\iii}^-~,
    \label{UDW-mon}
\end{align}
is a monopole operator, with
$$
\sigma_{\iii}^+ = \ket{1_\iii}\bra{0_\iii}
\qquad
\sigma_{\iii}^- = \ket{0_\iii}\bra{1_\iii}
$$
the Pauli ladder operators for each detector.  The function 
$$
\lambda_\iii(t)=\lambda_\iii\,\chi\left(\frac{t}{T_\iii}\right)\leq 
\lambda_\iii
$$
is a time-dependent coupling 
function of strength $\lambda_\iii$ and 
width $T_\iii$ for each detector.  The  smearing functions $F_{\iii}(\bx)=F\left(\frac{\bx-\bx_{\iii}}{\sigma_\iii}\right)$  define the (finite) region of spacetime in which each detector interacts with the field, centered at
the  positions $\bx_\iii$ and having
spatial width $\sigma_\iii$

Henceforth we shall set all couplings,
gaps, and widths to be equal.
Writing $H_{int} = \hat H$,
the interaction-picture time evolution operator for any initial state is
\begin{align}
    \hat U = \mc T \left[\exp\left(\ii \int_{-\infty}^\infty \dd t \hat H (t)\right)\right]\, .
\end{align}
In the weak coupling regime, the evolution operator $\hat U$ can be approximated using the Dyson series. To second order in the coupling constants, one finds
\begin{align}
    &\hat U=\hat U^{(0)}
    +\hat U^{(1)}
    +\hat U^{(2)}
    +\mc{O}(\lambda_{\iii} ^3)~,
\end{align}
where
\begin{subequations}
\begin{align}
\begin{split}
    \hat U^{(0)}&:=\openone
\end{split}\\
\begin{split}
    \hat U^{(1)}&:=-\ii\int_{\infty}^{\infty} \dd t \hat H (t)
\end{split}\\
\begin{split}
    \hat U^{(2)}&:=-\int_{\infty}^{\infty} \!\dd t \int_{-\infty}^{t} \!\dd t'\, \hat H (t)\hat H (t')~.
\end{split}
\end{align}
\end{subequations}
Setting the initial state to be a tensor product of the respective ground states,
\begin{align}
	\hat \rho_0=\ket{0_{\aaa} }\bra{0_{\aaa} }
	\otimes\ket{0_\bbb }\bra{0_\bbb }
	\otimes\ket{\psi}\bra{\psi}~,
\end{align}
the time-evolved state will be  $\hat \rho_T= \hat U \hat \rho_0 \hat U{}^\dagger $. To second order, this is
\begin{align}
    \hat \rho_T=\hat \rho_0
    +\hat \rho^{(1,0)}
    +\hat \rho^{(0,1)}
    +\hat \rho^{(1,1)}
    +\hat \rho^{(2,0)}
    +\hat \rho^{(0,2)}
    +\mc{O}(\lambda_{\aaa} ^3) \, .
\end{align}

Now consider the reduced density matrix for the detectors, after interacting with the field. The trace is a linear operation; we can apply it separately at every order in perturbation theory. Therefore
\begin{align}
    \rho_{\aaa\bbb}=
    &\ket{0_{\aaa} }\bra{0_{\aaa} }
	\otimes\ket{0_\bbb }\bra{0_\bbb }
    +\textrm{Tr}_\phi(\hat \rho^{(1,0)})
    \notag
    \\
    &+\textrm{Tr}_\phi(\hat \rho^{(0,1)})
    +\textrm{Tr}_\phi(\hat \rho^{(1,1)})
    \\
    &
    \notag
    +\textrm{Tr}_\phi(\hat \rho^{(2,0)})
    +\textrm{Tr}_\phi(\hat \rho^{(0,2)})
    +\mc{O}(\lambda_{\aaa} ^3) \, .
\end{align}
The first order terms $\textrm{Tr}_\phi(\hat \rho^{(0,1)})$ and $\textrm{Tr}_\phi(\hat \rho^{(1,0)})$ are dependent on the one-point correlator of the field, which for the vacuum is zero. This results in leading order corrections to the state of the detector being found at second order, which depend on the two-point correlation function of the field,
\begin{align}
     W(t,\bx,t',\bx ') &=
				\bra{\psi}
					\phi(t,\bx)
					\phi(t',\bx')
				\ket{\psi}\, ,\label{eq:Methods:TwoPoint-linear-dispersive}
\end{align}
also known as the Wightman function. In $(2+1)$ dimensions, the field expansion
\begin{equation}
    \hat\phi(t,\bx)=\!
    \int \!
    \frac{\dd^2 \bk  }{2\pi} 
    \frac{(\hat a_{\bk } u_k(t,\bx)
        +\hat a_{\bk }^\dagger u_k^*(t,\bx) 
    )}{\sqrt{2\omega_{\bk }}} \, ,
\label{field mode expansion}
\end{equation}
leads to a vacuum Wightman function
\begin{align}
    W(t,\bx,t',\bx ')=
    \int \!\dd^2
    \bk  \frac{e^{-\ii \omega_{\bk }  (t-t')+i \bk \cdot (\bx -\bx ')}
    }{2 (2 \pi )^2 \omega_{\bk } } \, .
\label{WGeneral}
\end{align}
The dependence of the reduced state on dispersion is contained entirely in this quantity.

Expressed as a bipartite density matrix for the detectors, the reduced density matrix in the basis $\{ \ket{0_{\aaa} }	\otimes\ket{0_\bbb }, \ket{1_{\aaa} }	\otimes\ket{0_\bbb }, \ket{0_{\aaa} }	\otimes\ket{1_\bbb }, \ket{1_{\aaa} }	\otimes\ket{1_\bbb } \}$ is given to leading order as
\begin{align}
    	\rho_{\aaa \bbb }=
    	\begin{pmatrix}
    		1-\mc L_{\aaa \aaa }-\mc L_{\bbb \bbb } &	0	&	0	&	\mc M	\\
    		0  &	\mc L_{\aaa \aaa }	&	\mc L_{\aaa \bbb }	&	0	\\
    		0  &	\mc L_{\bbb \aaa }	&	\mc L_{\bbb \bbb }	&	0	\\
    		\mc M^* &	0	&	0	&	0	\\
    	\end{pmatrix}
    	\, ,
    	\label{eq:Methods:DensityMatrix}
\end{align}
with the elements of \eqref{eq:Methods:DensityMatrix} given by 
\begin{align}
    \mathcal{L}_{\iii\jjj}  &=\int\dd^2\bk\frac{L_{\iii}(\bk)\left(L_{\jjj}(\bk)\right)^*}{2 (2\pi)^2\omega_{\bk}} \\
    \mathcal{M}           &=\int\dd^2\bk\frac{M(\bk)}{2 (2\pi)^2\omega_{\bk}}
\end{align}
where $(\iii,\jjj) \in (\aaa,\bbb)$ and 
\begin{widetext}
\begin{align}
    L_{\iii}(\bk) &= \lambda  {e^{\ii \bk \cdot \bx_\iii}}\widetilde F(\sigma |\bk|) \int _{-\infty}^\infty\!\!\!\dd t\, \chi(\tfrac{t}{T})\, e^{\ii\left(\Omega+\omega_{\bk}\right)t}
\\
    M(\bk) &=-2\lambda^2 e^{\ii \bk \cdot(\bx_\aaa-\bx_\bbb)} \left[\widetilde F(\sigma |\bk|)\right]^2 \int_{-\infty}^{\infty} \!\!\!\!\!\!\dd t \int_{-\infty}^{t}\!\!\!\!\!\!\dd t' e^{\ii \Omega (t+t')}e^{-\ii\omega_{\bk}(t-t')}\chi(\tfrac{t}{T})\,\chi(\tfrac{t'}{T})\, .
\end{align}



We will proceed in a similar way as \cite{pozas15aa} (see equations (23) and (24) of \cite{pozas15aa}, which are for 3+1 dimensional UDW detectors with Gaussian switching and smearing that do not take into account dispersion). Specializing to Gaussian switching,
\begin{align}
    \chi(\tfrac{t}{T})=e^{-\frac{t^2}{2 T^2}}\, ,
\end{align}
we find
\begin{align}
    L_{\iii}(\bk)&=
        \lambda 
        \widetilde F(\sigma|\bk|)
        e^{-\ii \bk \cdot \bx_\mu}
        G_1(\bk)
\\
    M(\bk)&=
        -\lambda^2
       \widetilde F(\sigma|\bk|)^2
        e^{\ii \bk \cdot(\bx_\aaa-\bx_\bbb)}
        G_2(\bk)
\end{align}
where $G_i$ are defined as
\begin{align}
G_1(\bk):=
    \int _{-\infty}^\infty\!\!\!\dd t\,
        e^{-\tfrac{t^2}{2 T^2}}\,
        e^{\ii\left(\Omega+\omega_{\bk}\right)t}= T \sqrt{2\pi}e^{-\frac{T^2}{2} \left(\Omega+\omega_{\bk}\right)^2}
\end{align}
and
\begin{align}
G_2(\bk)&:=
    2\int_{-\infty}^{\infty} \!\!\!\!\!\!\dd t \,
    \int_{-\infty}^{t}\!\!\!\!\!\!\dd t' 
    e^{\ii\Omega(t+t')}e^{-\ii\omega_{\bk}(t-t')}
    e^{-\tfrac{t^2}{2 T^2}}
        e^{-\tfrac{(t')^2}{2 T^2}}\, .
\end{align}
The function $G_2$ can be evaluated using techniques outlined in   appendix A of \cite{pozas15aa}; we obtain
\begin{align}
G_2(\bk)= 2\pi T^2 
    e^{-{T^2} \left(\Omega^2+\omega_{\bk}^2\right)}
   \textrm{erfc}
  \left(\ii \omega_k T\right) 
\end{align}
where
$\textrm{erfi}(z) = 1- \textrm{erf}(z)$ is the complementary error function.

Fortunately, in $(2+1)$ dimensions the angular part of the integral over $\bk$ is rather simple - it depends only on the imaginary exponential of the detector separation. Thus we calculate
\begin{align}
    \int_0^{2\pi}\dd\theta\, e^{-\ii |\bk||\bx_\mu|\cos\theta}
    =2 \pi  J_0(|\bk||\bx_\mu|),
\end{align}
where $J_0$ a Bessel function of the first kind. Hence
\begin{align}
    \mathcal{M}           &= -\lambda^2 \int_0^\infty \dd|\bk|\,\frac{\widetilde F(\sigma|\bk|)^2\,J_0(|\bk||\bx_\aaa-\bx_\bbb|)\,G_2(\bk)}{2 (2\pi) \omega_{\bk}}\\
    \mathcal{L}_{\aaa\bbb} &=   \lambda^2 \int_0^\infty \dd |\bk|\frac{|\bk|\,\widetilde F(\sigma|\bk|)^2\,J_0(|\bk||\bx_\aaa-\bx_\bbb|) G_1(\bk)^2}{2 (2\pi) \omega_{\bk}} \\
    \mathcal{L}_{\aaa\aaa} &= \mathcal{L}_{\bbb\bbb} :=\mathcal{L} =   \lambda^2 \int_0^\infty \dd |\bk|\frac{|\bk|\,\widetilde F(\sigma|\bk|)^2\,G_1(\bk)^2}{2 (2\pi) \omega_{\bk}}
\end{align}
Minor simplifications lead to the expressions in the main text.

\subsection{Continuous-detector Fields}

The basic idea of our proposal is to employ laser pulses as (continuous) UDW detectors (an idea proposed in 
\cite{gooding20aa}) 
with the surface fluctuations of the BEC playing the role of vacuum fluctuations of the scalar field, consistent with approaches in analog gravity.
To this end we must
  calculate the bipartite density matrix for the reduced continuous-detector system explicitly, for the case of identical switching functions and pointlike detectors. 

As discussed in the main text, perturbations $\delta \hat{E}_\jjj$ about the laser probe coherent amplitudes $E_{\jjj0}$ couple to BEC fluctuations according to equation $(4)$,
\begin{align}
     \hat{H}_\jjj(t)=-\alpha E_{\jjj0}(t,0)\delta \hat{E}_\jjj(t,0)\hat{\phi}(t,\bx_\jjj)\, .
\end{align}
We will use the same notation as in the two-level case, but with $\hat{\rho}_0=\hat{\rho}_{0,\aaa\bbb}\otimes |0\rangle \langle 0 |$, where 
  $|0\rangle$ is the $\hat{\phi}$ vacuum. The total interaction Hamiltonian is 
\begin{align}
    \hat{H}(t)=\sum_{\iii=\aaa,\bbb}\epsilon_\iii(t)  \hat{\mathcal{E}}_\iii(t)\hat{\phi}_\iii(t)\, ,
\end{align}
where $\varepsilon_\iii(t)=-\alpha E_{\iii0}(t,0)$ is the effective switching function identified in the main body and $\hat{\phi}_\iii(t)\equiv \hat{\phi}(t,\bm{x}_\iii)$ is evaluated at the $\iii^{th}$ interaction point (with $\iii\in \{\aaa,\bbb\}$). The quantity $\hat{\mathcal{E}}_\iii(t)\equiv\delta\hat{E}_\iii(t,0)$ is the EM field perturbation for the $\iii^{th}$ laser probe, polarized in the $x$-direction and evaluated on the BEC plane. In the Coulomb gauge, one finds $\delta\hat{E}_\iii(t,z=0)=-\partial_t \delta \hat{A}_\iii(t,z=0)$, where $\hat{A}_\iii(t,z)$ is the EM potential for the $\iii^{th}$ probe.

We denote the initial state of the detectors and field by $\ket{\Psi}=\ket{00}\otimes\ket{0}$, and the state long after the interaction by $\ket{\Psi_f}=\hat{U}\ket{\Psi}=\sum_n \ket{\Psi_f^{(n)}}$, with $\ket{\Psi_f^{(n)}}=\hat{U}^{(n)}\ket{\Psi}$. 
As we are interested in the final state of the lasers and not the BEC,
the reduced density operator is
\begin{align}
\hat{\rho}_{\aaa\bbb}=&\text{Tr}_\phi\left(\ket{\Psi_f} \bra{\Psi_f}\right) 
    = \sum_{n,m}\int d\mu\, \bra{\mu}\left(\ket{\Psi_f^{(n)}}\bra{\Psi_f^{(m)}}\right)\ket{\mu} 
    \equiv  \sum_{n,m}\hat{\rho}_{\aaa\bbb}^{(n,m)}\, ,
\end{align}
where $\ket{\mu}$ is an element of the Fock basis for the Hilbert space associated with $\hat{\phi}$. Explicitly, $\ket{\Psi_f^{(0)}}=\ket{\Psi}$, 
\begin{align}
    \ket{\Psi_f^{(1)}}=-i\int_{-\infty}^\infty dt\, \hat{H}(t)\ket{\Psi}\, ,
\end{align}
and
\begin{align}
    \ket{\Psi_f^{(2)}}=-\int_{-\infty}^\infty dt\,\int_{-\infty}^t dt'\, \hat{H}(t)\hat{H}(t')\ket{\Psi}\, .
\end{align}
Rearranging and making use of the completeness relation $\int d\mu\, |\mu\rangle \langle \mu |=1$, one finds the second-order reduced density operator perturbations
\begin{align}
\hat{\rho}_{\aaa\bbb}^{(1,1)}=\sum_{\iii,\jjj}\int_{-\infty}^{\infty}dt\, \int_{-\infty}^{\infty}dt'\, \varepsilon(t)\varepsilon(t')W_{\iii\jjj}(t,t')\hat{\mathcal{E}}_\iii(t')|00\rangle \langle 00|\hat{\mathcal{E}}_\jjj(t)
\, ,
\end{align}
 \begin{align}
\hat{\rho}_{\aaa\bbb}^{(2,0)}=-\sum_{\iii,\jjj}\int_{-\infty}^{\infty}dt\, \int_{-\infty}^{t}dt'\, \varepsilon(t)\varepsilon(t')W_{\iii\jjj}(t,t')\hat{\mathcal{E}}_\iii(t)\hat{\mathcal{E}}_\jjj(t')|00\rangle \langle 00|\, ,
\end{align}
and
\begin{align}
\hat{\rho}_{\aaa\bbb}^{(0,2)}=-\sum_{\iii,\jjj}\int_{-\infty}^{\infty}dt\, \int_{-\infty}^{t}dt'\, \varepsilon(t)\varepsilon(t')W_{\iii\jjj}(t,t')|00\rangle \langle 00|\hat{\mathcal{E}}_\jjj(t')\hat{\mathcal{E}}_\iii(t)\, ,
\end{align}
with $W_{\iii\jjj}(t,t')\equiv \langle 0|\hat{\phi}_\iii(t)\hat{\phi}_\jjj(t')|0\rangle$. 

The vacuum-vacuum matrix elements of the reduced density operator are $\bra{00}\hat{\rho}_{\aaa\bbb}^{(0,0)}\ket{00}=1$, $\bra{00}\hat{\rho}_{\aaa\bbb}^{(1,1)}\ket{00}=0$, and
\begin{align}
        \bra{00}\hat{\rho}_{\aaa\bbb}^{(2,0)}\ket{00}=&-\int_{-\infty}^{\infty}dt\int_{-\infty}^t dt'\, \varepsilon(t)\varepsilon(t') W_\aaa(t,t')\bra{00}\hat{\mathcal{E}}_\aaa(t)\hat{\mathcal{E}}_\aaa(t')\ket{00}\nonumber\\
        &-\int_{-\infty}^{\infty}dt\int_{-\infty}^t dt'\, \varepsilon(t)\varepsilon(t') W_\bbb(t,t')\bra{00}\hat{\mathcal{E}}_\bbb(t)\hat{\mathcal{E}}_\bbb(t')\ket{00}\, ,
\end{align}
where $W_\iii(t,t')=\bra{0}\hat{\phi}_\iii(t)\hat{\phi}_\iii(t')\ket{0}$. The only other contribution is given by $\bra{00}\hat{\rho}_{\aaa\bbb}^{(0,2)}\ket{00}=( \bra{00}\hat{\rho}_{\aaa\bbb}^{(2,0)}\ket{00})^*$.
 
The momentum states $\{ \ket{1_K} \}$ span the space of 
one-particle states in a $(1+1)$ scalar quantum field theory and obey  the completeness relation
\begin{align}\label{complete}
    1 = \int \frac{dK}{2\pi(2\Omega_K)}\ket{1_K}\bra{1_K}\, ,
\end{align}
which follows from the scalar product
\begin{align}
    \langle 1_K |1_{K'}\rangle=2\pi (2\Omega_K)\delta(K-K')\, .
\end{align}
Including states that are vacuum with respect to either detector field, the leading-order bipartite reduced state space has projector
\begin{align}
\hat{1}=&\ket{00}\bra{00}+\int\frac{dK}{2\pi(2\Omega_K)}\left(\ket{1_K 0}\bra{1_K 0}+\ket{0 1_K}\bra{0 1_K}\right)\nonumber\\ &+\int\frac{dK}{2\pi(2\Omega_K)}\int\frac{dK'}{2\pi(2\Omega_{K'})}\ket{1_K 1_{K'}}\bra{1_K 1_{K'}}\, , \label{projector}
\end{align}
where we have neglected double excitations of individual detectors, which for each pair of identical excitations (i.e. $K=K'$) do not contribute to the leading-order negativity \cite{PhysRevD.101.125020}. 
Inserting the projector \eqref{projector} on both sides of the reduced density operator $\hat{\rho}_{\aaa\bbb}$, one can obtain all relevant elements of the reduced density matrix, in the basis $\{ \ket{0_{\aaa} }	\otimes\ket{0_\bbb }, \ket{1_{\aaa,K} }	\otimes\ket{0_\bbb }, \ket{0_{\aaa} }	\otimes\ket{1_{\bbb,K} }, \ket{1_{\aaa,K} }	\otimes\ket{1_{\bbb,K} } \}$. For a fixed $K=K'$, this reduced density matrix can be expressed to leading order as
\begin{align}
    	\rho_{\aaa \bbb }=
    	\begin{pmatrix}
    		1-\mc L_{\aaa \aaa}-\mc L_{ \bbb\bbb } &	0	&	0	&	\mc M	\\
    		0  &	\mc L_{\aaa\aaa }	&	\mc L_{\aaa \bbb }	&	0	\\
    		0  &	\mc L_{\bbb \aaa }	&	\mc L_{\bbb\bbb }	&	0	\\
    		\mc M^* &	0	&	0	&	0	\\
    	\end{pmatrix}
    	\, ,
    	\label{eq:Methods:DensityMatrix2}
    \end{align}
just as in Appendix A. The elements are given by
\begin{align}
    \mc L_{\aaa\aaa } =\int_{-\infty}^{\infty}dt\, \int_{-\infty}^{\infty}dt'\,\varepsilon(t)\varepsilon(t')W_{\aaa\aaa}(t,t')\langle 00| \hat{\mathcal{E}}_{\aaa}(t)\hat{\mathcal{E}}_{\aaa}(t')|00\rangle\, ,
\end{align}
\begin{align}
    \mc L_{\aaa \bbb } =\int_{-\infty}^{\infty}dt\, \int_{-\infty}^{\infty}dt'\,\varepsilon(t)\varepsilon(t')W_{\aaa\bbb}(t,t')\langle 1_{\aaa,K}0|\hat{\mathcal{E}}_{\aaa}(t')|00\rangle \langle 00|\hat{\mathcal{E}}_{\bbb}(t)|01_{\bbb,K} \rangle\, ,
\end{align}
and 
\begin{align}
    \mc M =-2\int_{-\infty}^{\infty}dt\, \int_{-\infty}^{t}dt'\,\varepsilon(t)\varepsilon(t')W_{\aaa\bbb}(t',t)\langle 00| \hat{\mathcal{E}}_{\aaa}(t')\hat{\mathcal{E}}_{\bbb}(t)| 1_K 1_K\rangle\, .
\end{align}
The element $\mc L_{\bbb\bbb}$ is defined analogously to $\mc L_{\aaa\aaa}$, and it seems that for identical detectors, $\mc L_{\aaa \bbb }=\mc L_{\bbb \aaa }$, which follows from $W_{\aaa\bbb}(t,t')=W_{\bbb\aaa}(t,t')$ (along with $\mc L_{\aaa\aaa}=\mc L_{\bbb\bbb}$, which follows from $W_{\aaa\bbb}(t,t')=W_{\bbb\aaa}(t,t')$).

The reduced density matrix elements can be expressed more explicitly using mode expansions for the detector fields. Taking seriously the interpretation of $\delta\hat{A}_\iii(t,z)$ as a $(1+1)$-dimensional Klein-Gordon field perturbation, one has 
\begin{align}
\delta\hat{A}_\iii(t,z)=\int\frac{dK}{\sqrt{2\pi(2\Omega_K)}}\left(\hat{b}_{\iii K}e^{-i\Omega_K (t-z)}+\hat{b}_{\iii K}^\dagger e^{i\Omega_K (t-z)}\right)\, ,
\end{align}
which implies
\begin{align}
\hat{\mathcal{E}}_\iii(t)=-\partial_t \delta\hat{A}_\iii(t,0)  =-i\int dK\,\sqrt{\frac{\Omega_K}{4\pi}}\left(\hat{b}_{\iii K}e^{-i\Omega_K (t-z)}-\hat{b}_{\iii K}^\dagger e^{i\Omega_K (t-z)}\right)\, .
\end{align}
Using \eqref{complete}, the corresponding vacuum-vacuum two-point function of the $\varepsilon$ operators for the individual detector beams is
\begin{align}
    \bra{00}\hat{\mathcal{E}}_\iii(t)\hat{\mathcal{E}}_\iii(t')\ket{00}=\int dK\, \Omega_K^2\,e^{-i\Omega_K (t-t')}\, ,
\end{align}
regularized by the natural experimental UV cutoff defined by the laser spot size, which dictates the sensitivity band of the laser probes.

Transitions to excited detector states are determined by $\bra{1_{\iii K}}\hat{\mathcal{E}}_\iii(t)\ket{0_\iii}=2i\,e^{i\Omega_K t}\sqrt{\pi\Omega_K^3}$. The excited-excited matrix elements $\bra{1_K 1_{K'}}\hat{\rho}_{\aaa\bbb}^{(1,1)}\ket{1_K 1_{K'}}$ and $\bra{1_K 1_{K'}}\hat{\rho}_{\aaa\bbb}^{(2,0)}\ket{1_K 1_{K'}}$ both vanish at this order. The element $\bra{1_K 1_{K'}}\hat{\rho}_{\aaa\bbb}\ket{00}$ has components $\bra{1_K 1_{K'}}\hat{\rho}_{\aaa\bbb}^{(1,1)}\ket{00}=0$ and
\begin{align}
        \bra{1_K 1_{K'}}\hat{\rho}_{\aaa\bbb}^{(2,0)}\ket{00}=&4\pi\sqrt{(\Omega_K \Omega_{K'})^3}\int_{-\infty}^{\infty}dt\int_{-\infty}^t dt'\, \varepsilon(t)\varepsilon(t') W_{\aaa\bbb}(t,t')e^{i(\Omega_K t+\Omega_{K'}t')}\nonumber\\
        &+4\pi\sqrt{(\Omega_K \Omega_{K'})^3}\int_{-\infty}^{\infty}dt\int_{-\infty}^t dt'\, \varepsilon(t)\varepsilon(t') W_{\bbb\aaa}(t,t')e^{i(\Omega_K t'+\Omega_{K'}t)}\, ,
\end{align}
with $W_{\aaa\bbb}(t,t')=\bra{0}\hat{\phi}_\aaa(t)\hat{\phi}_\bbb(t')\ket{0}$. Furthermore we find 
$\bra{1_K 0}\hat{\rho}_{\aaa\bbb}^{(2,0)}\ket{1_{K'} 0}=0$, $\bra{0 1_K }\hat{\rho}_{\aaa\bbb}^{(2,0)}\ket{0 1_{K'}}=0$,  \begin{align}
\bra{1_K 0}\hat{\rho}_{\aaa\bbb}^{(1,1)}\ket{1_{K'} 0}=4\pi\sqrt{(\Omega_K \Omega_{K'})^3}\int_{-\infty}^{\infty}dt\int_{-\infty}^\infty dt'\, \varepsilon(t)\varepsilon(t') W_{\aaa}(t,t')e^{i(\Omega_K t'-\Omega_{K'}t)}\, ,
\end{align}
and
\begin{align}
        \bra{0 1_K }\hat{\rho}_{\aaa\bbb}^{(1,1)}\ket{0 1_{K'}}=4\pi\sqrt{(\Omega_K \Omega_{K'})^3}\int_{-\infty}^{\infty}dt\int_{-\infty}^\infty dt'\, \varepsilon(t)\varepsilon(t') W_{\bbb}(t,t')e^{i(\Omega_K t'-\Omega_{K'}t)}\, 
\end{align}
for the diagonal elements. 
The remaining off-diagonal terms are $\bra{1_K 0 }\hat{\rho}_{\aaa\bbb}^{(2,0)}\ket{0 1_{K'}}=0$ and
\begin{align}
        \bra{1_K 0 }\hat{\rho}_{\aaa\bbb}^{(1,1)}\ket{0 1_{K'}}=4\pi\sqrt{(\Omega_K \Omega_{K'})^3}\int_{-\infty}^{\infty}dt\int_{-\infty}^\infty dt'\, \varepsilon(t)\varepsilon(t') W_{\bbb\aaa}(t,t')e^{i(\Omega_K t'-\Omega_{K'}t)}\, .
\end{align}
For $K=K'$, each of these integrals has a counterpart of the same form in the analysis with a pair of identical two-level Unruh-DeWitt detectors. 



\section*{Appendix II: Experimental Setup}

\subsection{Modulated Detector Pulses}

The modulation bands in each beam lead to beating in the measured photocurrents at twice the modulation frequency. From the perspective of the Heisenberg picture, density fluctuations from the BEC are transduced into the modulated laser phases and carried along each beam at the beat frequency (taken to be in the MHz range). 
\end{widetext}

\begin{figure}[ht]
    \centering
    \includegraphics[width=0.85\columnwidth]{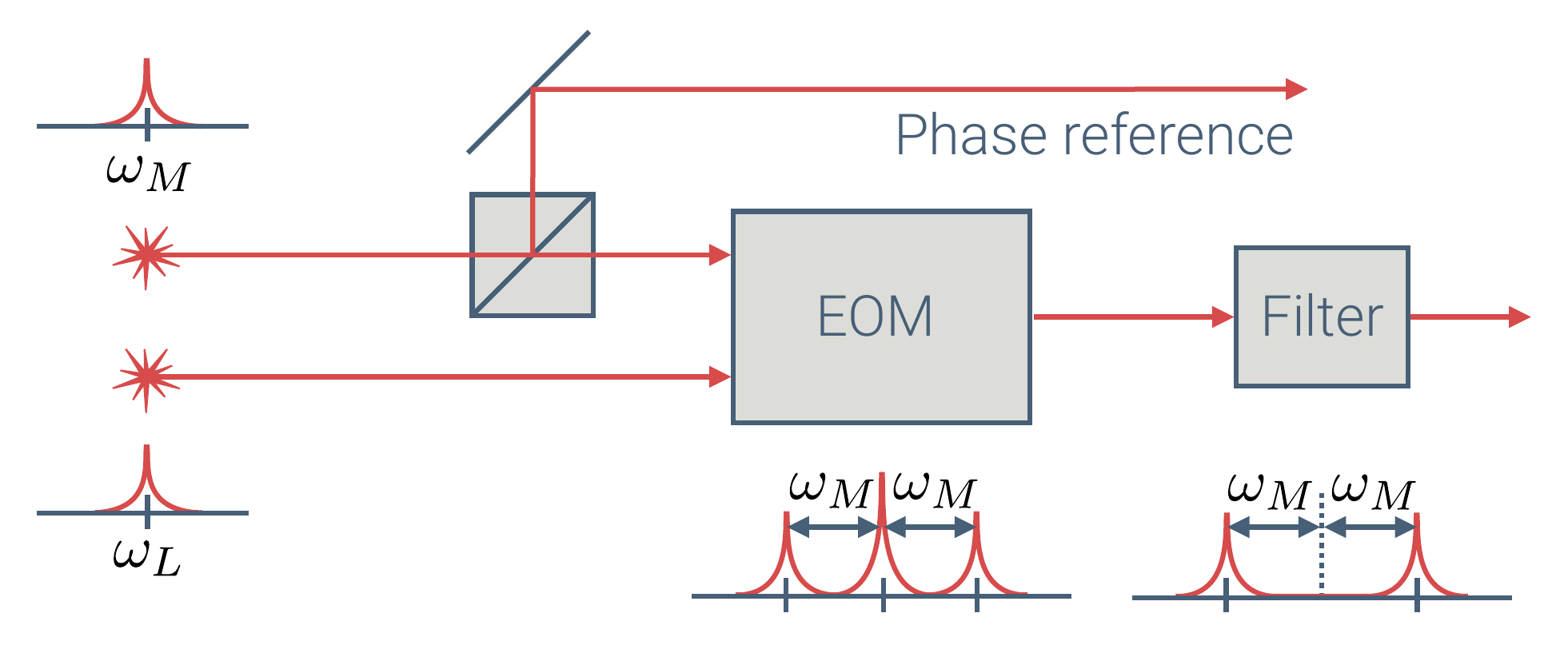}
    \caption{Individual detector beam preparation. Each initial laser beam is mixed with a microwave beam in an electro-optic modulator (EOM). The central peak is then filtered out of each modulated laser probe. A portion of the microwave beam is used as a phase reference for demodulation during the readout stage. Here, $\omega_M$ is the modulation frequency, and $\omega_L$ is the central frequency of the laser prior to modulation, which also corresponds to an atomic resonance frequency of the BEC species.}
    \label{fig:setup}
\end{figure}

By construction, the modulation bands in each beam receive opposite phase shifts from the BEC. Accordingly, the individual-beam difference-mode phase quadrature plays a prominent role, while the common-mode quadrature carries the majority of the excess laser noise. Let $\hat{Z}_{\jjj}(t)=(1/\sqrt{2})(\hat{b}_{\jjj+}(t)-\hat{b}_{\jjj-}(t))$ and $\hat{z}_{\jjj}(t)=(1/\sqrt{2})(\hat{b}_{\jjj+}(t)+\hat{b}_{\jjj-}(t))$ be the difference-mode and common-mode annihilation operators for the individual beams (respectively), with $\hat{b}_{\jjj\pm}$ being the upper ($+$) and lower ($-$) modulation band modes for the $\jjj^{th}$ beam. Fluctuations in the photon fluxes are given by
\begin{align}
\delta\tilde{n}_\jjj(t)= 4|\beta_{\jjj+}(t)|\cos^2\left(\omega_M t+\psi_\jjj\right)\, \hat{z}_\jjj^{\varphi_\jjj}(t)+2|\beta_{\jjj+}(t)|\sin 2\left(\omega_M t +\psi_\jjj \right)\, \tilde{\Pi}_\jjj^{\varphi_\jjj}(t)\,\label{Intfluc}
\end{align}
where $\tilde{\Pi}_\jjj^{\varphi_\jjj}(t)$ is the rotated momentum quadrature built from $\hat{Z}_{\jjj}(t)$, $\hat{z}_\jjj^{\varphi_\jjj}(t)$ is the rotated quadrature built from $\hat{z}_{\jjj}(t)$, $\beta_{\jjj+}$ are the (Gaussian) coherent amplitudes of our detector pulses, and $\{\varphi_\jjj\}$ are the individual beam phases, tunable through the individual local oscillator phases. The signal-carrying mode operators are denoted with a tilde and are given by
\begin{align}\label{Pitrans}
\tilde{\Pi}_\jjj^{\varphi_\jjj}(t)=\frac{1}{i\sqrt{2}}\left(e^{-i\varphi_\jjj}\tilde{Z}_\jjj(t)-e^{i\varphi_\jjj}\tilde{Z}_\jjj(t)^\dagger\right)
\end{align}
and $\tilde{Z}_\jjj(t)=\hat{Z}_\jjj(t)+\Delta\hat{Z}_\jjj(t)$, where $\Delta\hat{Z}_\jjj(t)$ decomposes into BEC density fluctuations and backaction noise.

Within this modulated detection scheme, it is natural to generalize the inseparability condition appearing in the main text using the finite-time spectral versions of the joint operators $(1/\sqrt{2})(\hat{z}_\aaa^{\varphi_\aaa}(t)\pm \hat{z}_\bbb^{\varphi_\bbb}(t))$ and $(1/\sqrt{2})(\hat{\Pi}_\aaa^{\varphi_\aaa}(t)\pm \hat{\Pi}_\bbb^{\varphi_\bbb}(t))$; in this case, the joint operators can be expressed as $\hat{q}_\pm(\Omega,T) = (1/\sqrt{2})(\hat{z}^{\varphi}_{\aaa}(\Omega,T)\pm \hat{z}^{\varphi}_{\bbb}(\Omega,T))$ and $\hat{p}_\pm(\Omega,T) = (1/\sqrt{2})(\hat{\Pi}_{\aaa}^{\varphi_\aaa}(\Omega,T)\pm \hat{\Pi}_{\bbb}^{\varphi_\bbb}(\Omega,T))$. The finite-time spectral modes are defined through windowed Fourier transforms of the form
\begin{equation}
    \hat{q}_{+}(\Omega,T)\equiv \frac{1}{\sqrt{T}}\int_{-T/2}^{T/2}dt\, e^{i\Omega t}\hat{q}_{+}(t)\, .
\end{equation}
To isolate these (unmodulated) modes, one can demodulate the two photocurrents, using the (tunable) phase references. Once demodulated, the pair of photocurrent signals provide direct access to the electromagnetic phase quadratures determined by $\hat{\mathcal{E}}_\aaa$ and $\hat{\mathcal{E}}_\bbb$.

\subsection{Generalised Inseparability}

For our task at hand, we seek expectation values of the photon flux \textit{after} the interaction has taken place; hence, in the interaction picture, the reduced density operator no longer evolves, and can be written simply as $\hat{\rho}_{\aaa\bbb}$:
\begin{align}
    \langle \hat{n}_\jjj(t)\rangle = \text{Tr}\left(\hat{\rho}_{\aaa\bbb}\hat{n}_\jjj(t)\right)\, .
\end{align}
For an arbitrary noise operator $\hat{O}$ in the joint detector Hilbert space, given the reduced density operator $\hat{\rho}_{\aaa\bbb}$, the variance is
\begin{equation}\label{redvar}
    V(\hat{O})\equiv \frac{1}{2}\text{Tr}\left(\hat{\rho}_{\aaa\bbb}\{ \hat{O}^\dagger,\hat{O} \}\right)\, .
\end{equation}
To second order, the reduced density operator is
\begin{align}
    \hat{\rho}_{\aaa\bbb}=\hat{\rho}_{0,\aaa\bbb}+\hat{\rho}_{\aaa\bbb}^{(1,1)}+\hat{\rho}_{\aaa\bbb}^{(2,0)}+\hat{\rho}_{\aaa\bbb}^{(0,2)}\, .
\end{align}
By linearity of the trace, the variance can then be written as
\begin{align}
    V(\hat{O})=V_0(\hat{O})+V^{(1,1)}(\hat{O})+V^{(2,0)}(\hat{O})+V^{(0,2)}(\hat{O}) \, .
\end{align}

Let us now focus on the generalised inseparability condition, which we express as
\begin{align}\label{DGCZ2}
\mc I(\Omega,T) = V(\hat{q}_+(\Omega,T)) + V(\hat{p}_-(\Omega,T))<1\, .
\end{align}
The first variance on the right-hand-side of \eqref{DGCZ2}
takes its vacuum value, since it depends only on common-mode operators, for which the BEC signal vanishes: $V(\hat{q}_+(\Omega,T'))=1/2$. For the second variance, one finds the expected leading behaviour $V_0(\hat{p}_-(\Omega,T'))=1/2$. Hence, the inseparability condition will be satisfied if
\begin{align}
    V^{(1,1)+(2,0)+(0,2)}(\hat{p}_-(\Omega,T'))<0\, ,
\end{align}
indicating entanglement. The main operator of experimental interest is therefore
\begin{align}
    \hat{p}_-(\Omega,T')=\frac{1}{\sqrt{T'}}\int_{T_0'}^{T_0'+T'}dt\, e^{i\Omega t}\, \hat{p}_-(t)\, ,
\end{align}
where
\begin{align}
    \hat{p}_-(t)=\frac{1}{\sqrt{2}}\left(\hat{\Pi}_\aaa^{\varphi_\aaa}(t)-\hat{\Pi}_\bbb^{\varphi_\bbb}(t)\right)
\end{align}
and the initial detection time $T_0'$ is sufficiently later than the interaction interval (i.e. $T_0'\gg T$). The $\Pi$ operators appearing in the EPR-like variable $\hat{p}_-(t)$ are the signal carriers for the analogue Unruh proposal \cite{gooding20aa}, for which the feasibility of detection has been argued. Though further analysis is required to firmly establish experimental feasibility of an entanglement measurement based on the inseparability condition~\eqref{DGCZ2}, as mentioned in the main body, there is experimental precedence for a time-domain version of the condition to be applicable for multimode laser fields, such as our pulsed laser probes~\cite{Zhang2015}. Hence, we find it to be an attractive candidate for an implementation of the entanglement harvesting protocol.

\bibliography{MDRbib}

\end{document}